\documentclass[twocolumn,printnumbers,amsmath,amssymb]{revtex4}
\usepackage{graphicx}% Include figure files
\usepackage{color}

\begin{document}

\title{Non-monotonic pressure dependence of the dynamics of soft glass-formers at high compressions}

\author{Lijin Wang}
\author{Yiheng Duan}
\author{Ning Xu$^*$}

\affiliation{CAS Key Laboratory of Soft Matter Chemistry, Hefei National Laboratory for Physical Sciences at the Microscale $\&$ Department of Physics, University of Science and Technology of China, Hefei 230026, People's Republic of China.}

\date{\today}

\begin{abstract}
In molecular dynamics simulations of soft glass-formers interacting via repulsions, we find that the glass transition temperature, dynamical heterogeneity, and glass fragility reach their maxima at the same crossover pressure $P_d$.  Our analysis of the zero-temperature jammed states indicates that states at $P_d$ have the highest bond orientational order with the largest spatial fluctuation.  Correspondingly, the low-frequency normal modes of vibration are the least localized and the average potential energy barrier along these modes are the highest for jammed states in the vicinity of $P_d$.  The reentrant glass transition and dynamics of supercooled liquids are thus predictable by these structural and vibrational precursors in the zero-temperature jammed states.
\end{abstract}

%\pacs{64.70.pv,63.50.Lm,61.43.Fs}

\maketitle

\section{Introduction}

Upon fast cooling or compression, a liquid or colloidal suspension undergoes the glass transition with sluggish dynamics.  Due to the lack of convincing phase transition signatures, the nature of the glass transition is still controversial \cite{debenedetti,angell,berthier1,parisi,gotze,lubchenko,elmatad,lerner,xu1,tanaka1}.  In practice, the glass transition temperature is usually defined as the temperature at which the relaxation time exceeds the measurable time window or is extrapolated from the divergence of a functional fit, {\it e.g.}~the  Vogel-Fulcher-Tamman (VFT) \cite{VFT} or mode coupling \cite{gotze} form, to the relaxation time data.  Previous studies have shown that the glass transition temperature usually increases monotonically with the pressure or equivalently the volume fraction, so does the glass fragility \cite{sastry,berthier2,zhang}.  Only very recently, it has been noticed that such a picture may be violated by core-softened colloids at high pressures.  For instance, a recent simulation has shown that the mode coupling glass transition temperature of a highly compressed model glass-former is reentrant as a function of the volume fraction \cite{berthier3}.

Core-softened colloids can exhibit complicated and rich phase behaviors and dynamics at high pressures \cite{berthier3,ikeda,osterman,camp,zhao,miller}, which have been mostly attributed to the softness of the colloids.  However, the underlying mechanism is apparently lacking.  Furthermore, our understanding of the dynamics of soft glass-formers at high compressions is still poor.  To our knowledge, the reentrant glass transition temperature \cite{berthier3} is the only phenomenon ever reported in the literature.  Other interesting and important aspects of soft glass-formers at high pressures, {\it e.g.}~dynamical heterogeneity and glass fragility, have not yet been concerned about.  Approaching the glass transition, the dynamics of supercooled liquids are spatially heterogeneous \cite{tanaka1,berthier1,lacevic,glotzer,weeks,berthier4,sausset}.  The dynamical heterogeneity tends to grow near the glass transition, implying a diverging dynamic correlation length.  As one of the most striking dynamical anomalies, will the dynamical heterogeneity show non-monotonic pressure dependence as well?

Near the glass transition a supercooled liquid loses ergodicity, caged in local potential energy basins over a long time before being able to escape.  From this point of view, the slow dynamics and dynamical heterogeneity of supercooled liquids are determined by the local fluctuation of the structural softness and the quasi-localized nature of the low-frequency normal modes of vibration of the metastable glass states at the local potential energy minima \cite{widmer-cooper,shintani,manning,chen1,tan}.  Due to the structural disorder, the low frequency modes of the metastable glasses are quasi-localized, {\it i.e.}~all particles are involved in the vibration but a small fraction in localized regions vibrate more strongly \cite{tan,schober,xu2,chen2}.  The quasi-localization is a special feature of the low-frequency vibrations of disordered solids and different from the Anderson localization in which most of the particles stay still except for a localized region.  The low-frequency vibrations are usually localized in soft spots, {\it i.e.}~more disordered regions of loosely constrained particles \cite{shintani,manning,chen1,tan}.

A recent study of the zero-temperature ($T=0$) marginally jammed solids has shown that along more quasi-localized low-frequency modes the potential energy barrier heights are lower \cite{xu2}.  Here jamming is restricted to packings of frictionless spheres interacting via repulsions at $T=0$.  A packing is jammed when it becomes rigid with nonzero elastic moduli and a coordination number larger than the isostatic value, {\it i.e.}~the minimum number of constraints per particle to maintain global mechanical stability \cite{ohern}.  The jammed states are actually metastable glasses with purely repulsive interactions.  The observed correlation between the energy barrier height and quasi-localization of jammed states suggests that a metastable glass with more quasi-localized low-frequency vibrations would be more vulnerable to excitations and consequently have a lower glass transition temperature.  To our knowledge, however, this picture has not yet been quantitatively demonstrated.

The study reported in this paper is inspired by our recent observation that at $T=0$ there exists a crossover volume fraction $\phi_d$ (or correspondingly a crossover pressure $P_d$) that separates marginally jammed solids from deeply jammed ones with distinct properties \cite{zhao}.  Our analysis indicates that the low-frequency quasi-localization gets weaker with increasing the volume fraction for marginally jammed solids until $\phi_d$ is attained, after which the quasi-localization of deeply jammed solids exhibits opposite volume fraction dependence \cite{zhao}.  From the picture discussed above, the reentrant quasi-localization at $\phi_d$ must have nontrivial effects on the glass transition and dynamics of supercooled liquids.

In this paper, we study the dynamics and structure of supercooled liquids consisting of soft particles in a wide range of pressures via molecular dynamics (MD) simulations.  We find that not only the glass transition temperature but also the dynamical heterogeneity and glass fragility are all reentrant near the crossover pressure $P_d$.  We propose an explanation of this non-monotonic pressure dependence from the analysis of the corresponding $T=0$ jammed states.  Our study provides quantitative evidence of the direct link between the glass transition temperature and low-frequency quasi-localization of the normal modes of the $T=0$ metastable states.

\section{Simulation details}
\label{sec:method}

Our model systems are three dimensional cubic boxes with periodic boundary conditions filled with a binary mixture of $N=1000$ frictionless spheres with the same mass $m$.  We also study some $N=10000$ systems and do not observe significant finite size effects.  The diameter ratio of the large to small spheres is $1.4$ to effectively avoid crystallization.  The potential between two particles $i$ and $j$ is $V\left(r_{ij}\right)=\epsilon \left( 1-r_{ij} / \sigma_{ij} \right)^{\alpha}/{\alpha}$ when their separation $r_{ij}$ is smaller than the sum of their radii $\sigma_{ij}$, and zero otherwise.  $\alpha$ is a tunable parameter to determine the softness of the interaction.  The same model system has been widely applied to the study of jamming and glass dynamics.  We study both harmonic ($\alpha=2$) and Hertzian ($\alpha=5/2$) repulsions.  In this paper, we only present the results for harmonic repulsion.  Our major findings are valid for Hertzian repulsion as well.  We set the characteristic energy scale $\epsilon$, small particle diameter $\sigma$, and particle mass $m$ to be the units.  The time and temperature are in the units of $\sigma/\sqrt{\epsilon/m}$ and $\epsilon/k_B$, respectively, where $k_B$ is the Boltzmann constant.

For glass-formers at $T>0$, we perform MD simulations at constant temperature and pressure to obtain the time evolution of particle motion according to the equations of motion \cite{allen}:
\begin{eqnarray}
\frac{{\rm d} \vec{r}_i}{{\rm d}t}&=&\vec{v}_i+\lambda\vec{r}_i,\\
\frac{{\rm d}\vec{v}_i}{{\rm d}t}&=&\frac{1}{m}\sum_{j\ne i}\vec{F}_{ij} - (\lambda + \zeta)\vec{v}_i,\\
\frac{{\rm d}L}{{\rm d}t}&=&L\lambda,
\end{eqnarray}
where $\vec{r}_i$ and $\vec{v}_i$ are the position and velocity of particle $i$, $\vec{F}_{ij}=-\nabla V_{ij}$ is the force acting on particle $i$ by particle $j$, $L$ is the length of the simulation box, and $\zeta$ and $\lambda$ are Lagrange multipliers to maintain constant temperature and pressure.  We apply 4-variable Gear predictor-corrector algorithm to integrate the equations numerically \cite{allen}.

The relaxation time of the liquids is measured from the self-part of the intermediate scattering function
\begin{equation}
F_{s}(k,t)=\frac{2}{N}\sum_{j} {\rm exp}({\rm i}\vec{k}\cdot[\vec{r}_{j}(t)-\vec{r}_{j}(0)]),
\end{equation}
where the sum is over all large particles, $\vec{r}_{j}(t)$ is the location of particle $j$ at time $t$, and $\vec{k}$ is chosen in the $x-$direction with $k=|\vec{k}|$ satisfying the periodic boundary conditions and being approximately the value at the first peak of the static structure factor.  The relaxation time $\tau$ is determined by $F_{s}(k,\tau)=e^{-1}F_{s}(k,0)$.  All the measures at $T>0$ discussed in this paper are taken after the system has been equilibrated for several $\tau$.

Widely accepted tools to probe the dynamical heterogeneity of supercooled liquids include the non-Gaussian parameter $\alpha_2$ \cite{weeks,yamamoto,rahman} and four-point dynamical susceptibility $\chi_4$ \cite{berthier1,lacevic,glotzer}.  The non-Gaussian parameter is defined as
\begin{equation}
\alpha_{2}(t)=\frac{\left<\Delta\vec{r}(t)^{4}\right>}{(1+2/d)\left<\Delta\vec{r}(t)^{2}\right>^{2}}-1,
\end{equation}
where $\Delta\vec{r}(t)=\vec{r}(t)-\vec{r}(0)$ is the particle displacement at time $t$, $d$ is the dimension of space, and $\left<. \right>$ denotes the time average over all the particles.  For simple liquids satisfying the Gaussian distribution, $\alpha_{2}=0$.  For supercooled liquids, however, $\alpha_{2}$ is greater than zero and shows interesting time and temperature dependence \cite{weeks}.  The four-point dynamical susceptibility $\chi_4$ captures the fluctuation of the number of mobile particles \cite{berthier1,lacevic,glotzer}.  It measures the variation of an overlap function $Q(a,t)$:
\begin{equation}
\chi_{4}(a,t)=\frac{1}{N}[\left<Q(a,t)^{2}\right>-\left<Q(a,t)\right>^{2}],
\end{equation}
where $\left< .\right>$ denotes the time average.  The overlap function $Q(a,t)=\sum_{i=1}^{N}W_{a}(|\vec{r}_{i}(t)-\vec{r}_{i}(0)|)$ evaluates the similarity between two configuration snapshots separated by a time interval $t$, where $a$ is a preset length, and $W_{a}(x)=1$ if $x\leq a$ and zero otherwise.  Note that here we only consider the self part of the overlap function, which dominates the inter-particle parts and describes the dynamical heterogeneity well \cite{glotzer,lacevic}.  Both $\alpha_2$ and $\chi_4$ are larger if the dynamics are more heterogeneous.

The structure of the glass-formers is evaluated from the pair distribution function of large particles \cite{allen}
\begin{equation}
g(r)=\frac{L^3}{(N / 2)^2}\left< \sum_i\sum_{j\ne i}\delta(r-r_{ij})\right>,
\end{equation}
where the sums are over all the large particles and $\left< .\right>$ denotes the time average.

We obtain jammed configurations at $T=0$ by quickly quenching ideal gas states to local potential energy minima using L-BFGS method \cite{lbfgs}.  We tune the volume fraction or equivalently the particle size successively until a desired pressure is attained.  At each pressure, we generate over $1000$ distinct jammed states and take the average over them.

In the $T=0$ jammed states, the local bond orientational order of particle $i$ is defined as \cite{steinhardt,lechner}
\begin{equation}
Q_6(i)=\sqrt{\frac{4\pi}{13} \sum_{m=-6}^6 \left| Q_{6m}(i)\right|^2},
\end{equation}
where
\begin{equation}
Q_{6m}(i)=\frac{1}{N_b(i)}\sum_{j=1}^{N_b(i)}Y_{6m}({\vec r}_{ij})
\end{equation}
with $N_b(i)$ the number of nearest neighbors of particle $i$ determined by the Voronoi tessellation, and $Y_{6m}({\vec r}_{ij})$ the spherical harmonics.  We measure the average bond orientational order $\left< Q_6\right>$ and its spatial fluctuation $\delta Q_6 = \sqrt{\left< Q_6^2\right> - \left< Q_6\right>^2}$, where $\left< .\right>$ denotes the particle and configuration average.

We diagonalize the Hessian matrix of the $T=0$ jammed states using ARPACK \cite{arpack} to obtain the normal modes of vibration.  For each mode, we calculate its participation ratio
\begin{equation}
p(\omega_n) = \frac{\left(\sum_{i=1}^N |\vec{e}_{n,i}|^2\right)^2}{N\sum_{i=1}^{N} |\vec{e}_{n,i}|^4},
\end{equation}
where $\omega_n$ and $\vec{e}_{n,i}$ are the frequency of the $n^{th}$ mode and polarization vector of particle $i$ in the mode.  The participation ratio measures the extensiveness of a mode, {\it i.e.}~the fraction of particles effectively involved in the vibration.  More localized modes have smaller participation ratios and vice versa.  We also measure the correlation function of the polarization vectors in mode $n$ \cite{zhao}:
\begin{equation}
C_n(r) = \frac{\sum_{i=1}^N\sum_{j=i}^N \vec{e}_{n,i}\cdot \vec{e}_{n,j}\delta(r-r_{ij})}{\sum_{i=1}^N\sum_{j=i}^N \delta(r-r_{ij})}.
\end{equation}
The normalized correlation function $C_n^N(r)=C_n(r)/C_n(0)$ can be well fitted with
\begin{equation}
C_n^N(r)=C_{n0}{\rm exp}\left( -r/\xi\right) +\Delta, \label{eq:correlation}
\end{equation}
where $C_{n0}$ and $\Delta$ are fitting parameters \cite{zhao}.  A more localized mode has a smaller correlation length $\xi$.

We use $\vec{R}_0$ to denote a $T=0$ jammed state in the configurational space.  If we move the state along mode $n$ to $\vec{R}=\vec{R}_0+u\vec{e}_n$, the potential energy rises to $V(\vec{R})=V(\vec{R}_0)+\Delta V$.  If $|u|$ is small enough, the perturbed state $\vec{R}$ is still in the same potential energy basin with $\vec{R}_0$ and the energy minimization will lead state $\vec{R}$ back to $\vec{R}_0$.  We thus define the potential energy barrier height along mode $n$, $\Delta V_{\rm max}(\omega_n)$ as the maximum $\Delta V$ above which state $\vec{R}$ is no longer in the same basin with $\vec{R}_0$.  Apparently, more stable jammed states should have higher $\Delta V_{\rm max}$ and hence possibly higher glass transition temperatures.  In the following sections we will show the interesting correlations between the measure of $Q_6$, $\Delta V_{\rm max}$, $p$, $\xi$, and the glass transition temperature.

\section{Dynamics and structure of supercooled soft glass-formers}
\label{sec:dynamics}

%%%%%%%%%%%%%%%%%%%%%%%%%%%%%%%%%%%%%%%%%%%%%%%%%%%%%%%%%%%%
\begin{figure}
\center
\includegraphics[width=0.47\textwidth]{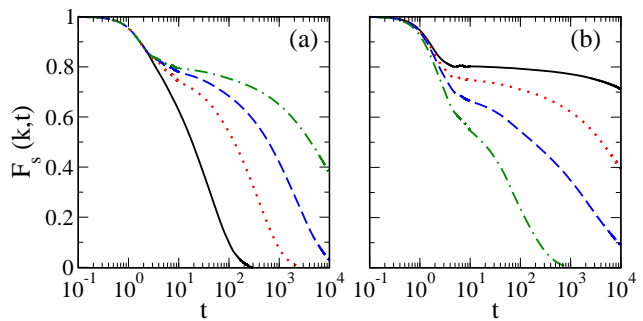}
\caption{\label{fig:fs} Self-part of the intermediate scattering function $F_{s}(k,t)$ with $\frac{kL}{2\pi}=10$ measured at $T=0.003$ and various pressures for systems with harmonic repulsion ($\alpha=2$). The pressures are $0.04$ (black solid), $0.07$ (red dotted), $0.08$ (blue dashed), and $0.09$ (green dot-dashed) in (a) and $0.2$ (black solid), $0.26$ (red dotted), $0.3$ (blue dashed), and $0.6$ (green dot-dashed) in (b).}
\end{figure}
%%%%%%%%%%%%%%%%%%%%%%%%%%%%%%%%%%%%%%%%%%%%%%%%%%%%%%%%%%%%

Fig.~\ref{fig:fs} shows the pressure evolution of the self-part of the intermediate scattering function $F_s(k,t)$ at a fixed temperature $T=0.003$ for systems with harmonic repulsion, which simulates the route of the colloidal glass transition.  As shown in Fig.~\ref{fig:fs}(a), the relaxation time increases rapidly with increasing the pressure as usually observed \cite{xu1,berthier2,brambilla}.  When the pressure gets higher, a typical two-step relaxation emerges due to the cage effect and becomes more and more pronounced.  From our previous experience, we may naturally expect that the glass transition happens at a critical pressure above which the systems turn into glasses, which is true for hard spheres \cite{brambilla} or soft spheres at low enough temperatures \cite{xu1,berthier2}.  At $T=0.003$, however, we surprisingly find that above a crossover pressure the relaxation time decreases with further compression.  As shown in Fig.~\ref{fig:fs}(b), the dynamics get faster at high compressions and the two-step relaxation eventually disappears at high enough pressures.

In order to have a quantitative picture of the pressure dependence of the glass transition, we measure the relaxation time in equilibrium at various temperatures and pressures.  At constant pressure, we fit the relaxation time with the VFT function
\begin{equation}
\tau = \tau_0 {\rm exp} \left( \frac{A}{T - T_0}\right), \label{vft}
\end{equation}
where $\tau_0$, $A$, and $T_0$ are fitting parameters.  $T_0$ is the VFT glass transition temperature at which $\tau=\infty$.

%%%%%%%%%%%%%%%%%%%%%%%%%%%%%%%%%%%%%%%%%%%%%%%%%%%%%%%%%%%%
\begin{figure}
%\vspace{0.1in}
\center
\includegraphics[width=0.47\textwidth]{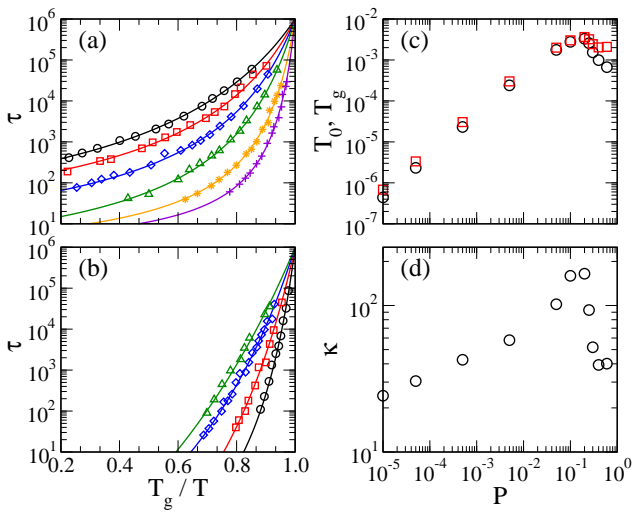}
\caption{\label{fig:angell} (a)$-$(b) Arrhenius plots of the relaxation time $\tau$, (c) VFT glass transition temperature $T_0$ at which $\tau$ diverges (black circles), glass transition temperature $T_g$ at which $\tau = 10^6$ (red squares), and (d) glass fragility $\kappa$ for systems with harmonic repulsion ($\alpha=2$).  The pressures are $0.00001$ (black circles), $0.00005$ (red squares), $0.0005$ (blue diamonds), $0.005$ (green triangles), $0.05$ (orange stars), and $0.1$ (violet pluses) in (a), and $0.2$ (black circles), $0.25$ (red squares), $0.3$ (blue diamonds), and $0.4$ (green triangles) in (b).  The solid curves are the fits with Eq.~(\ref{vft}). }
\end{figure}
%%%%%%%%%%%%%%%%%%%%%%%%%%%%%%%%%%%%%%%%%%%%%%%%%%%%%%%%%%%%

In panels (a) and (b) of Fig.~\ref{fig:angell} we plot the relaxation time measured at constant pressure as a function of $T_g/T$, {\it i.e.}~the Arrhenius plot, where $T_g$ is defined as the temperature at which $\tau=10^6$.  At all the pressures, our data can be well fitted with Eq.~(\ref{vft}), from which we can estimate $T_0$ and $T_g$.  As shown in Fig.~\ref{fig:angell}(c), both $T_0$ and $T_g$ vary non-monotonically with the pressure, in agreement with the recent observation that the mode coupling glass transition temperature is reentrant as a function of the volume fraction \cite{berthier3}.  $T_0$ and $T_g$ reach their maximum values ($\sim 0.003$) at a crossover pressure $P_d\approx 0.2$, which is approximately equal to the critical pressure separating marginal jamming from deep jamming at $T=0$ \cite{zhao}.  This agreement is not just an coincidence, which will be discussed in detail in Section~\ref{sec:jam}.

%%%%%%%%%%%%%%%%%%%%%%%%%%%%%%%%%%%%%%%%%%%%%%%%%%%%%%%%%%%%
\begin{figure}
\vspace{-0.08in}
%\center
\includegraphics[width=0.5\textwidth]{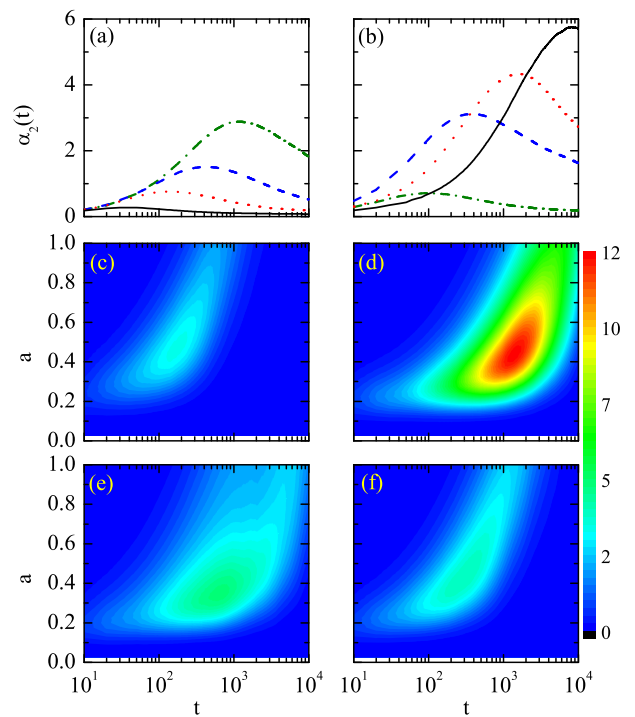}
\caption{\label{fig:d_h} Dynamical heterogeneity of supercooled liquids with harmonic repulsion ($\alpha=2$) measured at various pressures and a fixed temperature $T=0.003$.  Panels (a) and (b) show the non-Gaussian parameter $\alpha_2(t)$.  The pressures are $0.04$ (black solid), $0.07$ (red dotted), $0.08$ (blue dashed), and $0.09$ (green dot-dashed) in (a) and $0.2$ (black solid), $0.26$ (red dotted), $0.3$ (blue dashed), and $0.6$ (green dot-dashed) in (b).  Panels (c)$-$(f) are the contour plots of the four-point dynamical susceptibility $\chi_4(a,t)$ measured at pressures of (c) $0.06$, (d) $0.08$, (e) $0.4$, and (d) $0.6$.}
\end{figure}
%%%%%%%%%%%%%%%%%%%%%%%%%%%%%%%%%%%%%%%%%%%%%%%%%%%%%%%%%%%%

The slope of the Arrhenius plot at $T=T_g$, $\kappa=\frac{\partial ({\rm ln} \tau)}{\partial  (T_g/T)}|_{_{T=T_g}}$ measures the glass fragility.  A fragile (strong) glass has a large (small) $\kappa$.  Fig.~\ref{fig:angell}(a) shows that the glass fragility increases with the compression when the pressure $P<P_d$, in consistent with previous observations \cite{sastry,berthier2}.  However, this trend holds only up to $P\approx P_d$, as shown in Fig.~\ref{fig:angell}(b) that the glasses become stronger instead when $P>P_d$.  The explicit pressure dependence of $\kappa$ is plotted in Fig.~\ref{fig:angell}(d) with a maximum present at $P_d$, indicating that glasses at $P_d$ are the most fragile.  The gap between $T_0$ and $T_g$ also reflects the glass fragility to some extent, which can be seen from the expression of $\kappa$ assuming that the relaxation time satisfies Eq.~(\ref{vft}).  Strong glasses may have large values of $T_g-T_0$ due to their more Arrhenius behaviors, which is exactly the case in Fig.~\ref{fig:angell}(c).  It is interesting to know that high pressure has nontrivial effects on glass properties: it can melt glasses and make glasses stronger.  From panels (c) and (d) of Fig.~\ref{fig:angell}, we can see that with comparable fragility glasses at $P>P_d$ can have much higher glass transition temperature than those at $P<P_d$.  Our observation here thus proposes an effective way to obtain strong glasses with high glass transition temperatures.

Accompanied with the drastic slowdown upon the glass transition, the dynamics of supercooled liquids are spatially heterogeneous.  The dynamical heterogeneity grows with the increase of the relaxation time \cite{berthier1,lacevic,glotzer}.  Since the relaxation time is reentrant in pressure at constant temperature, is it possible that the dynamical heterogeneity exhibits the similar pressure dependence?  The answer is positive from Fig.~\ref{fig:d_h} which shows the non-Gaussian parameter $\alpha_2$ and four-point dynamical susceptibility $\chi_4$ measured at various pressures on both sides of $P_d$ and a fixed temperature $T=0.003$.

%%%%%%%%%%%%%%%%%%%%%%%%%%%%%%%%%%%%%%%%%%%%%%%%%%%%%%%%%%%%
\begin{figure}
\center
\includegraphics[width=0.47\textwidth]{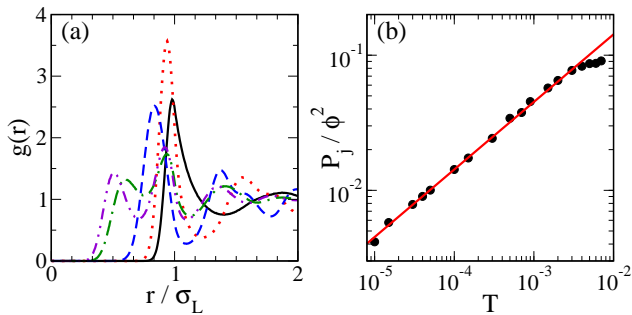}
\caption{\label{fig:gr} (a) Pair distribution function of large particles $g(r)$ of supercooled liquids with harmonic repulsion ($\alpha=2$) measured at a fixed temperature $T=0.003$ and pressures of $0.01$ (black solid), $0.06$ (red dotted), $0.2$ (blue dashed), $0.4$ (green dot-dashed), and $0.5$ (violet dot-dot-dashed).  (b) Crossover pressure $P_j$ at which the first peak of $g(r)$ reaches the maximum height at constant temperature.  The red solid line shows the scaling relation $P_j/\phi^2 \sim T^{1/2}$ which fits the data well up to the maximum glass transition temperature $T_g^{\rm max}\approx 0.003$.}
\end{figure}
%%%%%%%%%%%%%%%%%%%%%%%%%%%%%%%%%%%%%%%%%%%%%%%%%%%%%%%%%%%%

At all the pressures the non-Gaussian parameter $\alpha_2(t)$ is peaked at a time $t_{\alpha}$ scaled with the relaxation time $\tau$.  The magnitude of the peak signifies the strength of the dynamical heterogeneity.  Panels (a) and (b) of Fig.~\ref{fig:d_h} indicate that $\alpha_2$ behaves opposite pressure dependence on the two sides of $P_d$.  When $P<P_d$, the peak of $\alpha_2$ grows at an increscent $t_{\alpha}$, meaning the growth of the dynamical heterogeneity upon compression.  When $P>P_d$, however, the peak of $\alpha_2$ drops instead and moves to a shorter $t_{\alpha}$.

Panels (c)-(f) of Fig.~\ref{fig:d_h} are contour plots of the four-point dynamical susceptibility $\chi_4(a,t)$ measured at different pressures.  $\chi_4$ reaches its maximum at a length $a\approx 0.4$ weakly depending on the pressure and a time $t_{\chi}$ showing similar pressure dependence as $\tau$ and $t_{\alpha}$.  Along with the compression, the peak value of $\chi_4$ initially grows until $P\approx P_d$ and then drops, as what $\alpha_2$ does.  Like $\alpha_2$, large $\chi_4$ means strong dynamical heterogeneity.  Fig.~\ref{fig:d_h} thus indicates that at the crossover pressure $P_d$, together with the emergence of the maximum glass transition temperature and glass fragility, the dynamics are the most heterogeneous.

Our recent study of the $T=0$ jammed states suggests that the crossover pressure $P_d$ may have its structural origins \cite{zhao}.  At $P_d$, the coordination number of jammed states is $12$, implying that particles may start to interact with their second nearest neighbors.  Structurally a second peak thus emerges at $r<\sigma_L$ in the pair distribution function of large particles $g(r)$, where $\sigma_L$ is the diameter of large particles.  For soft glass formers studied here, we observe the similar structural change across $P_d$.  Fig.~\ref{fig:gr}(a) shows the pressure evolution of $g(r)$ at $T=0.003$.  Compared to the $T=0$ results \cite{zhao}, the thermal motion smears out the discontinuous jump of $g(r)$ at $r=\sigma_L$ and masks the initial formation of the second peak on the left hand side of $r=\sigma_L$ at $P\approx P_d$.  However, the emergence of the second peak (or even more peaks at extremely high pressures) at $r<\sigma_L$ is still a robust signature that distinguishes the glass-formers on the two sides of $P_d$.

It has been shown that during the formation of soft colloidal glasses at a fixed low temperature the first peak of $g(r)$ reaches the maximum height at a crossover pressure $P_j$, which is embedded in the glass regime and reminisces the $T=0$ jamming transition \cite{zhang,wang,berthier5}.  Beyond the maximum glass transition temperature $T_g^{\rm max}\sim 0.003$ (see Fig.~\ref{fig:angell}), because there are no solid states, $P_j$ is no longer associated with jamming.  As shown in Fig.~\ref{fig:gr}(b), at low temperatures up to $T_g^{\rm max}$, $P_j/\phi^2$ is scaled well with $T^{(\alpha-1)/\alpha}$ as reported \cite{zhang,wang}.  When $T>T_g^{\rm max}$, it is interesting that the scaling breaks down and $P_j$ remains almost constant in the temperature, corresponding to the decrease of the volume fraction \cite{jacquin}.

All the results discussed in this section indicate that the existence of the crossover at $P_d$ is not an accident, because all the quantities that we concern about undergo apparent changes there.   According to the proposal that the dynamics of supercooled liquids reflects the structural and vibrational properties of the $T=0$ metastable states \cite{widmer-cooper,shintani,manning,chen1,tan}, we may be able to seek the origin of the reentrant glass transition from the analysis of the $T=0$ jammed states.

\section{Implications from $T=0$ jammed states}
\label{sec:jam}

We recently find that there exists a crossover volume fraction $\phi_d\approx 1.2$ that divides the $T=0$ jammed states into marginally and deeply jammed regimes with distinct structures and power-law scaling of typical quantities such as the potential energy, elastic moduli, and coordination number \cite{zhao}.  The pressure at the crossover agrees with $P_d$.  We have also made a prediction that the non-monotonic pressure dependence of the quasi-localization of the low-frequency normal modes of vibration leads to the reentrant glass transition.  In this section, we will show that the unusual dynamics at high compressions discussed in section~\ref{sec:dynamics} are indeed strongly correlated to the properties of the $T=0$ jammed states.

%%%%%%%%%%%%%%%%%%%%%%%%%%%%%%%%%%%%%%%%%%%%%%%%%%%%%%%%%%%%
\begin{figure}
\vspace{-0.38in}
\center
\includegraphics[width=0.5\textwidth]{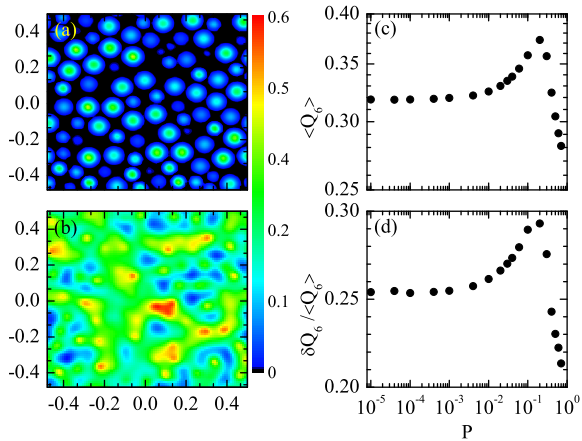}
\caption{\label{fig:op} Structure of the $T=0$ jammed states with harmonic repulsion ($\alpha=2$).  Panels (a) and (b) are the potential energy field in a two-dimensional cross section of the states at $P=0.00001$ and $0.5$, respectively.  Panels (c) and (d) show the pressure dependence of the average bond orientational order $\left< Q_6\right>$ and its relative fluctuation $\delta Q_6/\left< Q_6\right>$.}
\end{figure}
%%%%%%%%%%%%%%%%%%%%%%%%%%%%%%%%%%%%%%%%%%%%%%%%%%%%%%%%%%%%

Upon compression, each particle interacts with more and more neighbors.  It thus sounds plausible that the heterogeneity induced by the particle size dispersion may be suppressed and the jammed states may be more and more ordered in structure.  Panels (a) and (b) of Fig.~\ref{fig:op} show the potential energy distribution in a two-dimensional cross section of the jammed states on both sides of $P_d$, scanned by a point tracer.  When $P<P_d$, we can clearly see the boundary of each particle, indicating that particles only interact with their nearest neighbors.  When $P>P_d$, it is hard to identify single particles due to the penetration of the particle interactions to farther neighbors.  The potential energy field is still heterogeneous in space with no visible increase of the structural order.

We then show in Fig.~\ref{fig:op}(c) the bond-orientational order $\left<Q_6\right>$ of jammed states as a function of the pressure.  A state with a larger $\left<Q_6\right>$ is more ordered and vice versa.  When $P<P_d$ the jammed states are more ordered upon compression, as expected.  It is interesting that the order stops increasing at $P_d$.  When $P>P_d$, the compression boosts the disorder instead.  Recent studies have demonstrated that the particle displacement in a supercooled liquid is strongly correlated to the structural order of the particle \cite{shintani,tan}.  Disordered particles with small $Q_6$ are usually loosely constrained and tend to move to longer distances than ordered particles.  Therefore, disordered states should be more vulnerable than ordered states to the same strength of external excitations.  From this phenomenological picture, it is reasonable to expect a reentrant glass transition around $P_d$, which is exactly observed in our MD simulations.

%%%%%%%%%%%%%%%%%%%%%%%%%%%%%%%%%%%%%%%%%%%%%%%%%%%%%%%%%%%%
\begin{figure}
\center
\includegraphics[width=0.4\textwidth]{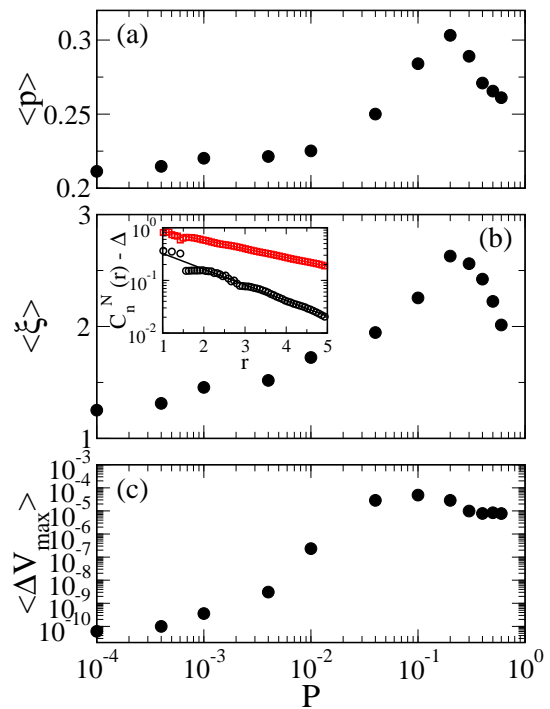}
\caption{\label{fig:ql} Pressure dependence of the participation ratio $\left< p\right>$, correlation length of the polarization vectors $\left<\xi \right>$, and potential energy barrier height along vibrational modes $\left<\Delta V_{\rm max} \right>$ averaged over the 20 lowest frequency normal modes of vibration for $T=0$ jammed states with harmonic repulsion ($\alpha=2$).  The inset to (b) shows the normalized correlation function $C_n^N(r)$ at $P=0.001$ (black circles) and $0.2$ (red squares).  The lines are the fits with Eq.~(\ref{eq:correlation}) from which $\xi$ is extracted.}
\end{figure}
%%%%%%%%%%%%%%%%%%%%%%%%%%%%%%%%%%%%%%%%%%%%%%%%%%%%%%%%%%%%

The fluctuation of the bond-orientational order, $\delta Q_6 /\left<Q_6\right>$ behaves similarly to $\left<Q_6\right>$, as shown in Fig.~\ref{fig:op}(d).  The fluctuation of the local structural order reaches the maximum at $P_d$.  In consideration of the correlation between the particle dynamics of supercooled liquids and structural order of metastable states, the reentrant dynamical heterogeneity in the vicinity of $P_d$ originates from the pressure dependence of the structural order fluctuation of the $T=0$ jammed states.

The normal modes of vibration are the fundamentals to understand the properties of solids under excitations, {\it e.g.}~the energy transport.  Recent studies have suggested that the low-frequency modes play important roles in the determination of the heterogeneous dynamics of supercooled liquids or sheared glasses \cite{widmer-cooper,shintani,manning,chen1,tan}.  As discussed above, disordered particles with small $Q_6$ are loosely constrained and easy to move.  In the low-frequency vibrations, these particles also form local soft spots with longer polarization vectors than the others and result in the quasi-localization.  We may then be able to find precursors of the reentrant glass transition in the analysis of the low-frequency modes.

Fig.~\ref{fig:ql}(a) shows the average participation ratio $\left< p\right>$ of the 20 lowest frequency modes.  $\left< p\right>$ reaches the maximum at $P_d$, meaning that the low-frequency modes at $P_d$ are the least localized, which is consistent with the presence of the maximum order.  To quantitatively illustrate the pressure dependence of the localization, we extract the correlation length $\xi$ from the correlation function of the polarization vectors $C_n(r)$ (see Section~\ref{sec:method} for details).  Interestingly, Fig.~\ref{fig:ql}(b) shows that $\xi$ is the largest at $P_d$ as well, which is another strong evidence of the least localization at $P_d$ \cite{zhao}.

In the potential energy landscape, each jammed state sits at the local potential energy minimum of a basin of attraction.  When thermally excited, the system explores the nearby configurational space.  Each basin has a potential energy barrier in any direction.  If the kinetic energy is large enough, the system will overcome the barriers and the glass melts.  In this picture, the glass transition temperature is correlated to the potential energy barrier heights.  In Fig.~\ref{fig:ql}(c) we show the potential energy barrier height $\left< \Delta V_{\rm max}\right>$ averaged over the 20 lowest frequency modes.  The jammed states around $P_d$ are the most stable with approximately the highest $\left< \Delta V_{\rm max}\right>$, although the peak of $\left< \Delta V_{\rm max}\right>$ is not exactly at $P_d$ probably due to some uncertainties in the calculation such as the large configurational variation of $\Delta V_{\rm max}$ and unphysical effects of the energy minimization algorithm.

From the analysis of the structural order and normal modes of vibration of the $T=0$ jammed states, we are able to understand the non-monotonic pressure dependence of the dynamics at high compressions.  Here we show the strong couplings between the structural order, quasi-localization, and energy barrier height.  All these quantities are reentrant around $P_d$, acting as the precursors of the reentrant glass transition.

\section{Conclusions}

%%%%%%%%%%%%%%%%%%%%%%%%%%%%%%%%%%%%%%%%%%%%%%%%%%%%%%%%%%%%
\begin{figure}
\center
\includegraphics[width=0.35\textwidth]{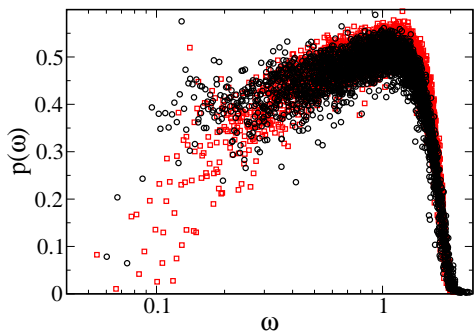}
\caption{\label{fig:lj} Participation ratio $p(\omega)$ of two three-dimensional $T=0$ metastable states consisting of $1000$ particles interacting via Lennard-Jones (black circles) and repulsive Lennard-Jones (red squares) interactions at a reduced number density $\rho=1.2$.  The systems are exactly the same as in Ref. 46.}
\end{figure}
%%%%%%%%%%%%%%%%%%%%%%%%%%%%%%%%%%%%%%%%%%%%%%%%%%%%%%%%%%%%

In this study, we find that soft glass-formers with purely repulsive interactions behave non-monotonic dynamics upon compression.  There exists a crossover pressure $P_d$ at which the glass transition temperature, glass fragility, and dynamical heterogeneity reach their maxima.  Although a large number of studies have been performed to understand the elusive glass transition problems, the reentrant glass transition and dynamics are still quite unexpected and may thus raise new challenging questions.  The unusual pressure dependence of the highly compressed glasses may have some potential applications.  For instance, we can produce strong glasses with high glass transition temperatures using highly compressed core-softened colloids.  Although strong glasses can be obtained at low compressions as well, the glass transition temperature has to be extremely low as the compensation (see Fig.~\ref{fig:angell}).

The results reported in this paper require experimental verification.  Possible candidates include extremely soft colloids such as star polymers, charged colloids with long range electrostatic interactions, or dusty plasmas.  Here we only concern about purely repulsive systems.  It is also interesting to know whether and how the attraction alters the picture.

The crossover pressure $P_d$ is consistent with that separating marginal jamming from deep jamming at $T=0$ \cite{zhao}, which inspires us to search for origins of the reentrant glass transition from the analysis of the $T=0$ jammed states.  Interestingly, we observe strong correlations between the structural order and properties of the normal modes of vibration of the $T=0$ jammed states.  The jammed states at $P_d$ are the most stable with the highest bond orientational order, the largest spatial fluctuation of the order, the least localized low-frequency modes, and the highest potential energy barriers.  It is thus not hard to understand why the glass transition is the highest and the dynamics are the most heterogeneous at $P_d$.

Due to the strong correlation between the glass transition temperature and properties of metastable glasses such as the structural order, low-frequency quasi-localization, and energy barrier height, we are able to predict and understand the effects of some conditional changes ({\it e.g.}~type of interactions and boundary conditions) on the glass transition from the analysis of the metastable states, {\it e.g.}~the participation ratio.  For instance, recent studies have shown that the inclusion of the attraction significantly influences the dynamics of supercooled liquids: A Lennard-Jones (LJ) system has a higher glass transition temperature than its counterpart with repulsive Lennard-Jones (RLJ) interactions \cite{berthier6,zhang2}.  This result is predictable from the participation ratio of the low-frequency vibrational modes of the metastable states.  The low-frequency modes of the RLJ system must be more localized than those of the LJ systems, which is exactly the case shown in Fig.~\ref{fig:lj}.  We believe that the participation ratio of low-frequency vibrational modes of metastable states is a simple but efficient tool to predict the behaviors of supercooled liquids in various systems.

This work is supported by National Basic Research Program of China (973 Program) No. 2012CB821500, National Natural Science Foundation of China No. 91027001 and 11074228, CAS 100-Talent Program No. 2030020004, and Fundamental Research Funds for the Central Universities No. 2340000034.

\end{document}